\documentclass[12pt, draftclsnofoot, onecolumn]{IEEEtran}
\usepackage{amsfonts}
\usepackage{amssymb}
\usepackage{amsthm}
\usepackage{amsmath,amsfonts,amssymb}
\usepackage[dvips]{graphicx}
\usepackage{verbatim}
\usepackage{bm}
\usepackage[ruled,vlined]{algorithm2e}
\usepackage{cite}
\usepackage{color}

\IEEEoverridecommandlockouts

\begin{document}

\title{Enabling UAV Cellular with Millimeter-Wave Communication: Potentials and Approaches}

\author{Zhenyu Xiao,~
        Pengfei Xia,~
and Xiang-Gen Xia
\thanks{Zhenyu Xiao is with the department of EIE, BKL-NCATM and BL-GAT of Beihang University.}
\thanks{Pengfei Xia is with Tongji University.}
\thanks{Xiang-Gen Xia is with University of Delaware.}}

\maketitle
\begin{abstract}
To support high data rate urgent or ad hoc communications, we consider mmWave UAV cellular networks and the associated challenges and solutions. To enable fast beamforming training and tracking, we first investigate a hierarchical structure of beamforming codebooks and design of hierarchical codebooks with different beam widths via the sub-array techniques. We next examine the Doppler effect as a result of UAV movement and find that the Doppler effect may not be catastrophic when high gain directional transmission is used. We further explore the use of millimeter wave spatial division multiple access and demonstrate its clear advantage in improving the cellular network capacity. We also explore different ways of dealing with signal blockage and point out that possible adaptive UAV cruising algorithms would be necessary to counteract signal blockage. Finally, we identify a close relationship between UAV positioning and directional millimeter wave user discovery, where update of the former may directly impact the latter and vice versa.
\end{abstract}


\section{Introduction}

\begin{figure}[t]
\begin{center}
  \includegraphics[width=12 cm]{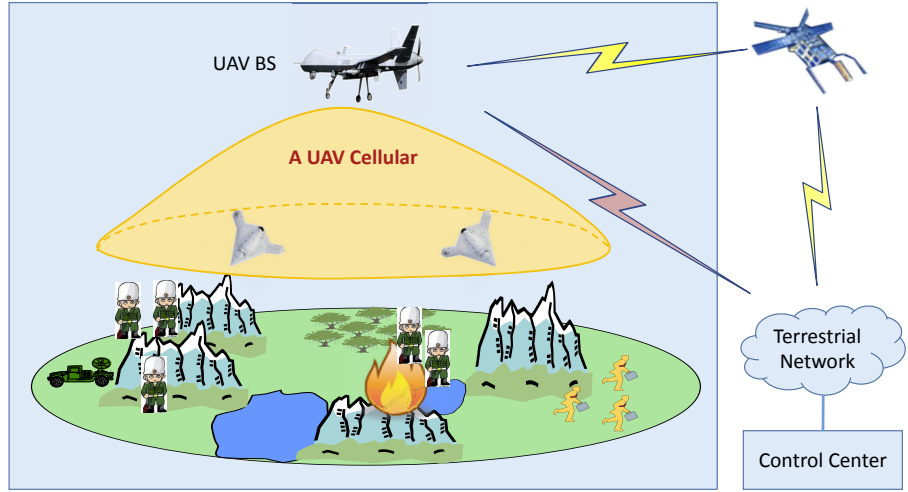}
  \caption{Illustration of a typical UAV cellular system.}
  \label{fig:system}\vspace{-0.1 in}
\end{center}
\end{figure}

Unmanned aerial vehicles (UAVs) have received increasing attention in the past decade~\cite{bekmezci2013flying,ponda2010decentralized}, thanks to  potential applications in reconnaissance, fire-fighting, aerial photo, remote sensing, disaster rescue and others. For the above scenarios where a fixed infrastructure network is destroyed or does not exist, it is important to quickly deploy a UAV cellular network to support urgent or ad hoc communications for the ground and low-altitude users.

A typical UAV cellular network is shown in Fig. \ref{fig:system}, where the base station (BS) is mounted on a flying UAV in the air, and mobile stations (MS) are distributed on the ground or in the low-altitude air. The UAV BS may be connected with the terrestrial networks via a satellite link or an air-to-ground wireless link. Typically, the traffic between MS and UAV BS includes circumstance information, control commands, and sensing data from various sensors, e.g. camera sensors~\cite{bekmezci2013flying,ponda2010decentralized}. In many cases where the large video monitoring traffic data from many camera sensors need to be collected and sent back to a control station for fast response, high data rate communication links between the MS and UAV BS are desirable. For this reason, we study in the paper millimeter wave communications for UAV cellular, as abundant frequency spectrum resource exists in the millimeter-wave (mmWave) frequency band~\cite{roh2014millimeter,rangan2014millimeter}.

The main difference between a mmWave UAV cellular network and a regular mmWave cellular network with fixed BS is that the UAV BS may move around.
Hence, the challenges of a regular mmWave cellular apply to the mmWave UAV cellular as well, including range and directional communications, rapid channel variation, multi-user access, blockage and others~\cite{rangan2014millimeter}. Some of the existing challenges are intensified due to UAV movement. For example, more efficient beamforming training and tracking are needed to account for UAV movement, and channel Doppler effect needs extra consideration. UAV movement also gives rise to some new challenges. For example, in mmWave UAV cellular networks, UAV position and user discovery are intertwined. On one hand, with a fixed position, UAV would be able to discover only the nearby users. On the other hand, UAV needs to find all potential users to serve to optimize its self positioning. Some of other existing challenges may actually be alleviated due to UAV mobility. For example, blockage is a significant performance limiting factor for regular mmWave cellular networks. In a mmWave UAV network, however, intelligent cruising algorithms may be developed to enable a UAV to fly out of a blockage zone and establish line of sight communications with an MS.

In this article, we investigate these key challenges in the mmWave UAV cellular and discuss possible solutions. Section II discusses mmWave wireless channel propagation characteristics, including link budget challenges and channel modeling. Section III presents design of a full hierarchy of codebooks to enable fast beamforming training and tracking for mmWave cellular networks. Section IV discusses the benefit of performing mmWave spatial division multiple access (SDMA) and illustrates the potential performance improvements. Section V addresses how to deal with blockage in mmWave UAV cellular networks and Section VI discusses the interaction between UAV directional user discovery and UAV positioning.

\section{Channel Propagation Characteristics}

When considering a mmWave UAV cellular, an immediate concern is the extremely high propagation loss, since Friis' transmission law states that the free space omnidirectional path loss grows with the square of the carrier frequency. Fortunately, the small wavelength of mmWave signals also enables greater (proportional to the square of the carrier frequency)  antenna gain for the same physical antenna size~\cite{rangan2014millimeter}. Consequently, higher carrier frequency does not in itself result in any increased propagation loss, provided that the antenna area remains fixed and suitable high gain antennas (and thus directional transmissions) are used at the BS. It is further shown in \cite{roh2014millimeter} that if the MS uses a directional antenna as well, the received power of the mmWave signals could be even higher than that of the low frequency signals. This indicates that mmWave wireless communication does not necessarily suffer from a link budget deficiency issue, and also demonstrates the importance of transceiver beamforming towards mmWave UAV cellular systems~\cite{rangan2014millimeter}.

High power consumption of mixed signal components as well as expensive radio-frequency (RF) chains makes it difficult, if not impossible, to realize full-blown digital baseband beamforming/precoding in mmWave communications. Instead, analog beamforming/hybrid precoding structures are usually preferred to support one or more stream transmissions \cite{sun2014mimo}, where all the antennas share a small number (much smaller than the number of antennas) of RF chains and generally have constant-amplitude beamforming/precoding coefficients \cite{sun2014mimo,alkhateeb2014channel}. Typically for the MS one stream transmission may be used, which consists of a single RF chain and $N_{\rm{MS}}$ antennas. For the BS to support multi-user communications, multi-stream transmission may be used, which has $N_{\rm{RF}}$ RF chains and $N_{\rm{BS}}$ antennas. Typically,  $N_{\rm{RF}}<N_{\rm{BS}}$.

MmWave channels are expected to have limited scattering \cite{rangan2014millimeter,alkhateeb2014channel} and multipath components (MPCs) are mainly generated by first- and second-order reflections, with different physical angles of departure (AoDs) and angles of arrival (AoAs). Since the number of MPCs is basically much smaller than the number of antennas, the AoDs and AoAs are sparse in the angle domain. The mmWave UAV cellular channel may share the same model as a regular mmWave cellular. However, unique to a mmWave UAV cellular system is that there is generally much less reflection around the UAV in the air than the reflection around the mobile user on the ground. 
Different MPCs have very close steering angles on the UAV side and may be grouped into a very small number of clusters, and the overall channel would be very sparse in the angle domain. As a result, compressive sensing based channel estimation approach, such as \cite{alkhateeb2014channel}, may be well suited especially for UAV mmWave systems. 
An uplink wide-band time-varying continuous channel model for a mmWave UAV cellular can be expressed as \cite{schniter2014channel}
\begin{equation}
{\bf{H}}(t) = \sqrt {N_{\rm{MS}}N_{\rm{BS}}} \sum\limits_{\ell  = 1}^{L(t)} {\lambda _\ell }(t)p(t-\tau_\ell(t)){\bf{a}}{({N_{\rm{BS}}},{\psi}(t))}{\bf{a}}({N_{\rm{MS}}},{\Omega _\ell }(t))^{\rm{H}},
\end{equation}
where $\lambda_\ell(t)$ is the complex coefficient of the $\ell$th path, $L(t)$ is the number of MPCs, $p(t)$ is the raised cosine pulse, $\tau_\ell(t)$ is the relative delay of the $\ell$th MPC, $\psi(t)$ is the AoA at the BS, while $\Omega _\ell(t)$ are the AoDs from the MS, ${\bf{a}}(\cdot)$ is the steering vector depending on the number of antennas and the steering angles. In general, only a very small number of strong MPCs may be searched out to form beams between BS and MS. As a result, the effect of delay spread may be further mitigated by spatial beamforming \cite{rangan2014millimeter}. Moreover, the channel coherence time is in fact relatively long versus the packet duration in mmWave communication (see Section III-B); thus the channel can usually be seen quasi-static. Hence, for simplicity a narrow-band discrete channel model
\cite{alkhateeb2014channel,sun2014mimo}:
\begin{equation} \label{eq_channel}
{\bf{H}} = \sqrt {N_{\rm{MS}}N_{\rm{BS}}} \sum\limits_{\ell  = 1}^{L} {\lambda _\ell }{\bf{a}}{({N_{\rm{BS}}},{\psi})}{\bf{a}}({N_{\rm{MS}}},{\Omega _\ell })^{\rm{H}}
\end{equation}
has also been extensively adopted.

\section{Fast Beamforming Training and Tracking}
In a mmWave UAV cellular, beamforming is required to steer along strong MPCs at both the BS and MS to provide necessary Tx/Rx antenna gains. Compared with conventional mmWave communications for static stations, the time constraint for beamforming training is more stringent due to UAV movement. Here we will discuss the challenges and promising solutions.

\subsection{Hierarchical Beam Search and Codebook Design}
Switched beamforming performs Tx/Rx joint beam search based on pre-designed codebooks. An exhaustive search algorithm, which sequentially tests all combinations of beam directions in the angle domain and finds the best pair of Tx/Rx beamforming codewords, is conceptually straightforward. Yet the overall search time is prohibitively costly due to the very large number of candidate directions.

\begin{figure}
\begin{minipage}[t]{0.6\linewidth}
\centering
\includegraphics[width=9.0 cm]{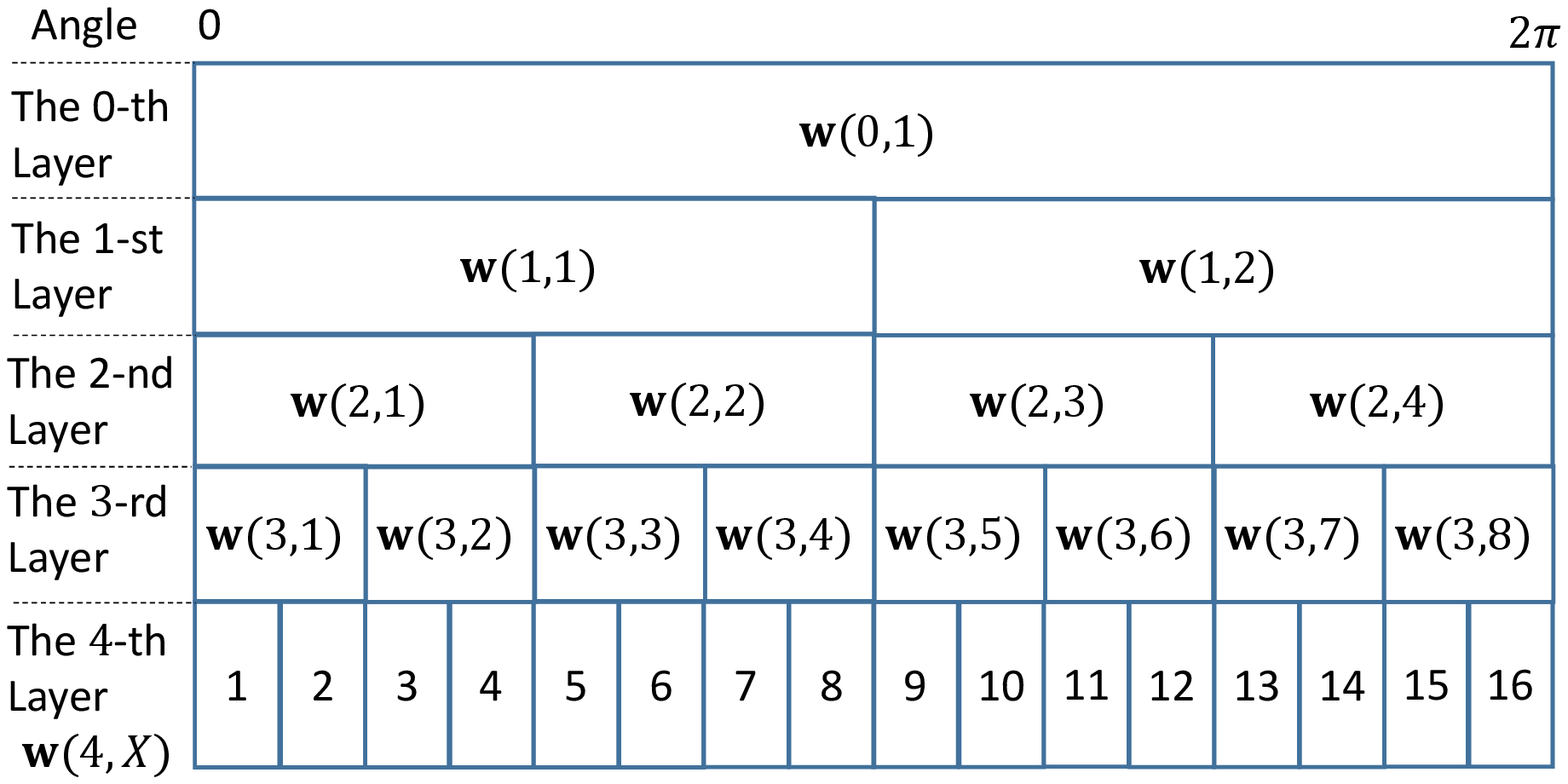}
\end{minipage}
\begin{minipage}[t]{0.4\linewidth}
\centering
\includegraphics[width=7.0 cm]{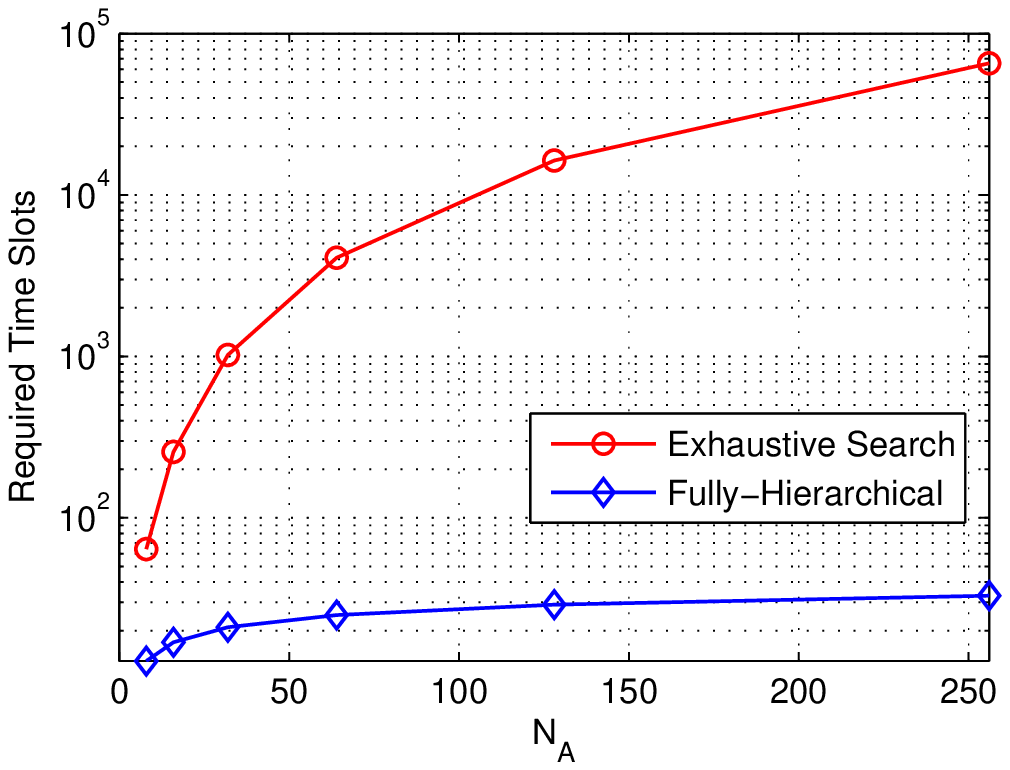}
\end{minipage}
\caption{\textbf{Left}: The beam coverage of the codebook for the fully-hierarchical scheme, where $N=16$ and $M=2$. \textbf{Right}: The comparison of time complexity between relevant search schemes, where the numbers of antennas at the BS and MS are the same ($N_{\rm{A}}$), $M=2$.} \vspace{-0.1 in}
\label{fig:Hier_Codebook_timecomp}
\end{figure}

To reduce the antenna training time and associated overhead, hierarchial beam search schemes based on a tree-structured beamforming codebook may be adopted~\cite{wang_2009_beam_codebook}. A typical hierarchial codebook $\mathcal{F}$ is shown in the left hand side of Fig. \ref{fig:Hier_Codebook_timecomp} with $N=16$ antennas  (larger antenna array may be needed in practice) and a degree of $M=2$. In the $k$th layer, there are $M^k$ codewords of the same beam width with different steering angles and collectively covering the entire search space in the angle domain. 
Let ${\bf{w}}(k,n)$ denote the $n$th codeword in the $k$th layer, $n=0,1,...,M^k$. Then the beam coverage of ${\bf{w}}(k,n)$ is approximately the union of the beam coverage of the $M$ codewords on the $(k+1)$-st layer $\{{\bf{w}}(k+1,(n-1)M+m)\}_{m=1,2,...,M}$.
Right hand side of Fig. \ref{fig:Hier_Codebook_timecomp} illustrates the training overhead comparison in terms of the required time slots between the fully hierarchical scheme and the exhaustive search scheme. It can be seen that the complexity of the fully hierarchical scheme is significantly lower than the exhaustive search scheme.

To enable hierarchical beam search, we need to design a full hierarchy of codebooks on all layers. The challenge is how to design codewords with wide beam width subject to the constant-amplitude (CA) constraint. It is even more challenging when the transmitter is constrained to have only one to two RF chains. 
A tree structured hierarchy of codebooks~\cite{he2015suboptimal} was designed using brute-force antenna deactivation (DEACT), where wider beams are generated by turning off part of the antennas. For mmWave wireless communications, separate power amplifiers for each antenna are usually employed to distribute the overall power amplification task across multiple independent amplifiers. For the DEACT approach, the total transmit power is usually small due to the small number of active antennas (and henceforth the small number of active power amplifiers).

Hybrid analog/digital beamforming/precoding was studied in~\cite{alkhateeb2014channel}, and the codebook design is formulated as a sparse compressive sensing problem and may be solved using e.g. orthogonal matching pursuit algorithms. Although multiple RF chains provide additional degrees of freedom, good wide-beam codewords may be generated only when the number of RF chains is large enough. When the number of RF chains is small, deep sinks \cite[Fig. 5]{alkhateeb2014channel} within the wide coverage have been observed, and degrade the overall training performance.

\begin{figure}
\begin{minipage}[t]{0.5\linewidth}
\centering
\includegraphics[width=9.0 cm]{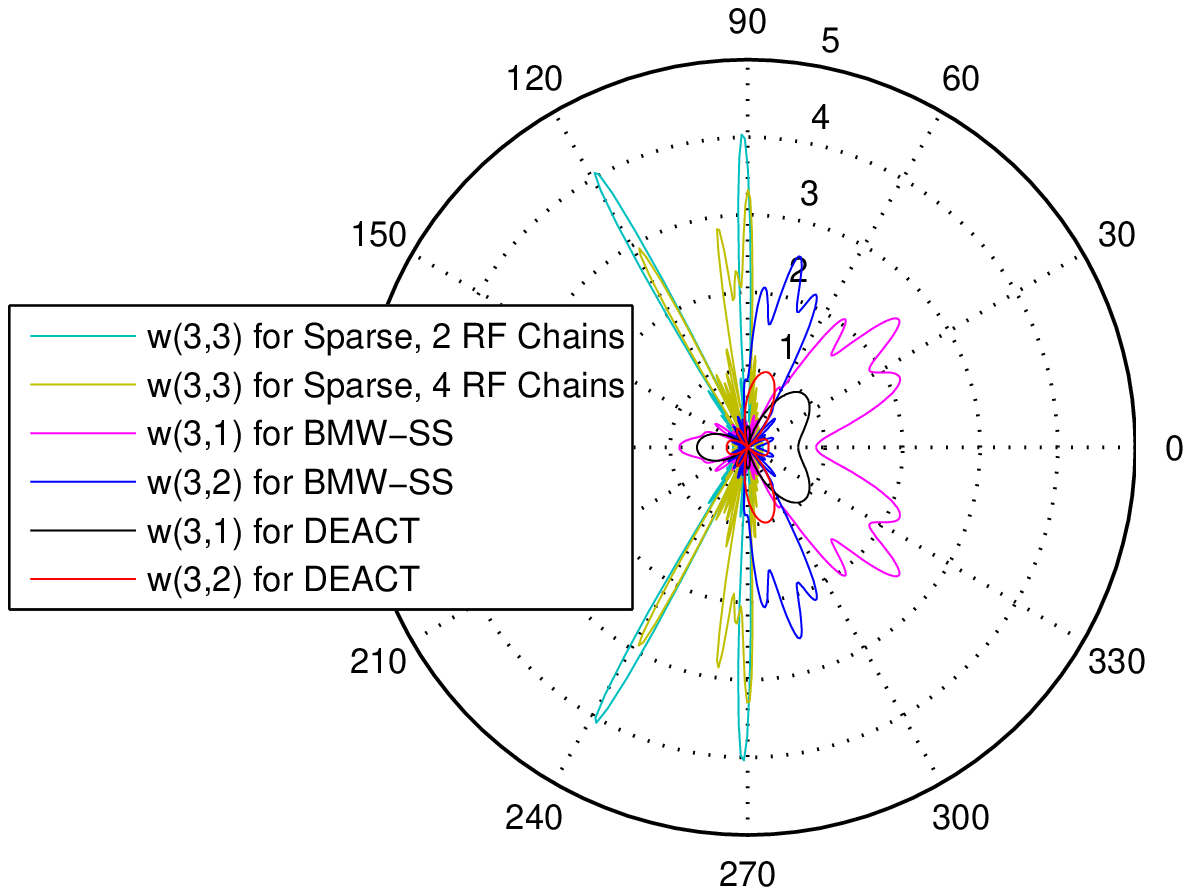}
\end{minipage}
\begin{minipage}[t]{0.5\linewidth}
\centering
\includegraphics[width=8.5 cm]{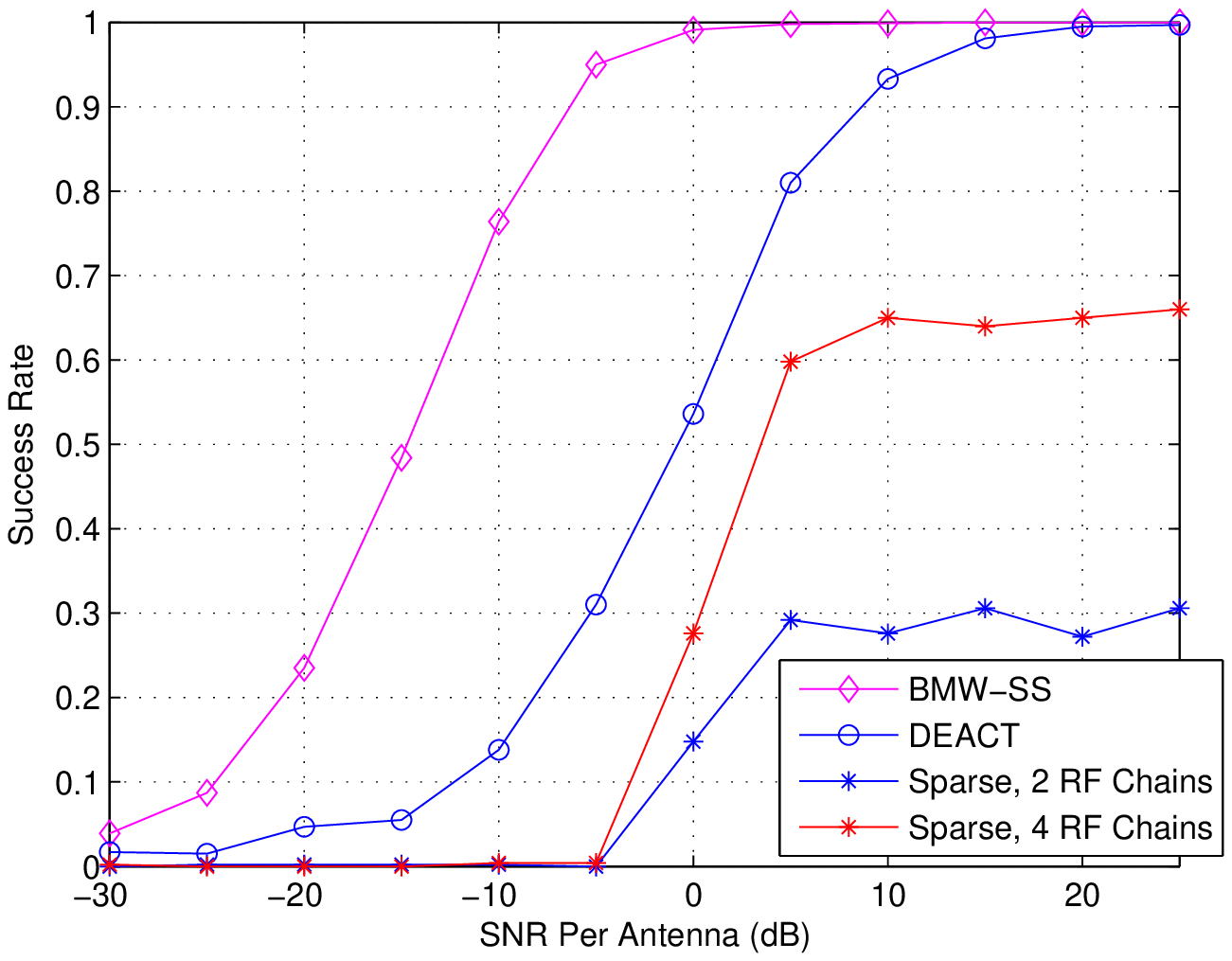}
\end{minipage}
\caption{\textbf{Left}: Comparison of the beam patterns, where $N=32$. $L_d=1$ for the Sparse approach in \cite{alkhateeb2014channel}. \textbf{Right}: Comparison of the success rates. $N_{\rm{BS}}=N_{\rm{MS}}=32$, $L=3$, and the power of the LOS component is 20 dB higher than the NLOS components.}
\label{fig:cmp_beampattern_success_rate}\vspace{-0.1 in}
\end{figure}

A full hierarchy of codebooks was recently designed and the BeaM Widening was achieved via a Single-RF Sub-array, and hence the name BMW-SS. A comparison of the beam patterns is shown on the left hand side of  Fig. \ref{fig:cmp_beampattern_success_rate}.
Compared with \cite{he2015suboptimal}, the BMW-SS approach is able to form very wide beams, but not by turning off part of antennas and the associated power amplifiers, and hence without sacrificing the total transmit power. On the other hand for the sparse codebooks in~\cite{alkhateeb2014channel}, when the number of RF chains is small, there exist clearly deep sinks within a wide beam coverage, and the sink is more severe when the number of RF chains is smaller (in accordance with \cite[Fig. 5]{alkhateeb2014channel}). In comparison,  the wide beams of BMW-SS are formed via only a single RF chain and does not suffer from deep sinks within a wide beam coverage.

The right hand side of Fig. \ref{fig:cmp_beampattern_success_rate} shows the comparison of the success (detection) rate, defined as the rate that the LOS component is successfully acquired in the beam search process. It is found that the BMW-SS approach achieves the best performance. Compared with DEACT, BMW-SS has a significant SNR gain due to the larger number of active antennas. Both BMW-SS and DEACT are able to achieve a success rate of 100\% in high SNR. However, the sparse codebooks in~\cite{alkhateeb2014channel} cannot achieve a success rate of 100\% even in high SNR, due to the deep sink within the beam coverage.

\subsection{Channel Variation and Beam Tracking}
As UAV itself may be moving, one might initially think that Doppler spread would be high and cause catastrophic effects to high rate transmissions. Suppose a UAV movement speed of $v=20$ m/s, a carrier wavelength of 5 millimeter, and an angle of $\theta = \pi/3$ for the angle between the moving direction and the UAV-MS linking direction. Conventionally, the channel coherence time may be approximated as $1/(v\cos(\theta)/\lambda) \approx 0.5$ milliseconds and the Doppler spread may be calculated as $f_{\rm{D}}=2$ KHz. This, however, may not be true in mmWave communications. As shown in~\cite{rangan2014millimeter}, the Doppler spread is actually a function of carrier frequency, mobile velocity, as well as the total angular dispersion, while the last term has not been taken into account in the conventional computations. According to the measurement results therein, mmWave signals generally arrive in a small number of path clusters, each with a relatively small angular spread. Moreover, directional transmission with narrow beams will further reduce the multipath angular spread. As a result, the individually resolvable MPCs  will vary slowly although the overall channel variation may be large. Similar observations are made in~\cite{Va2015Basic} that the realistic channel coherence time depends on the beam width and would be much larger when narrow beams are formed between the transmitter and receiver.


To further improve the training/tracking efficiency, \emph{a priori}  information regarding the distribution range of beamforming angles may be used. For example, in certain practices, the range of the steering angles may be only a subset of $[0,2\pi)$. Another example is that in certain practices, location of the MS may be available to the moving BS. Such \emph{a priori} information, together with the UAV movement information (such as GPS location, movement direction, speed) would help further reduce the beamforming training and tracking overhead. The hierarchical tree structure of the codebooks may also be used to enable fast tracking of steering beams. For instance, let ${\bf{w}}(S,i)$ be the beam direction acquired after a beam training process. The neighboring beam directions ${\bf{w}}(S,i-1)$ and ${\bf{w}}(S,i+1)$ may serve as short list candidates in the beam tracking process. Overall, many important topics remain open for UAV mmWave beamforming training and tracking.

\section{MMWave Spatial Division Multiple Access}

Due to the highly directional transmissions in mmWave, users from different directions may be well separated using different spatial beams. Hence, multiple users with different beams may access the channel at the same time. This is generally known as spatial division multiple access (SDMA), or beam division multiple access (BDMA)~\cite{Sun2015BDMA}. Theoretically speaking, when the BS is equipped with $N_{\rm{RF}}$ transceiver RF chains and each MS is equipped with a single transceiver RF chain, the overall multi-user capacity may be boosted by up to $N_{\rm{RF}}$ times when SDMA is used. 

\begin{figure}[t]
\begin{center}
  \includegraphics[width=7 cm]{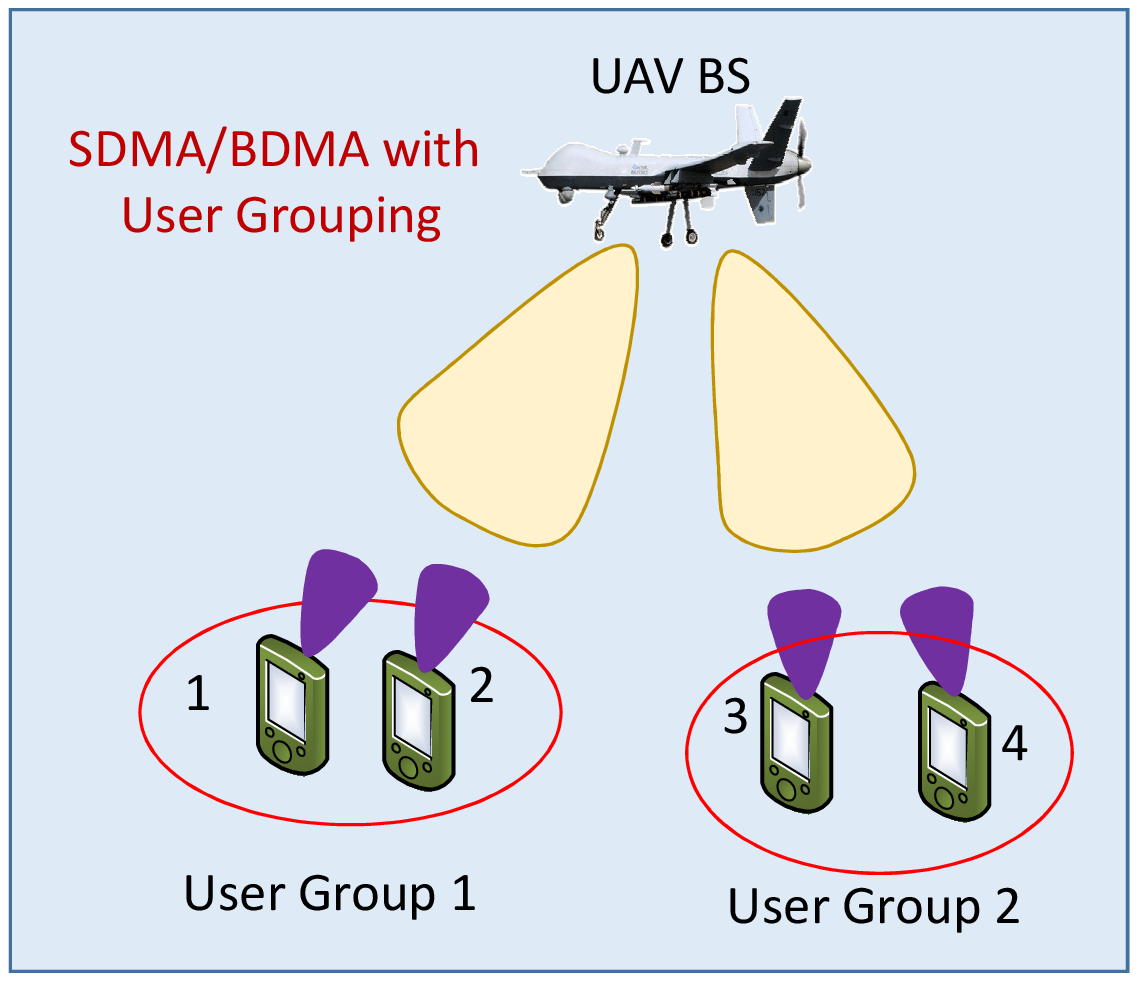}
  \caption{Only users from different groups may access the channel at the same time in mmWave SDMA. }
  \label{fig:BDMA_group}\vspace{-0.1 in}
\end{center}
\end{figure}

A critical issue of SDMA is how to group the users so that different users from different groups may access the BS at the same time, while not causing significant interference to each other. A simple yet practical strategy is to group users according to their AoDs, i.e., the steering angles at the BS side, and only users from different spatial groups are allowed to access the channel at the same time. In particular, it is possible to divide the entire range of AoDs $[0,2\pi)$ into $N_{\rm{BS}}$ clusters, while each cluster may be represented by a codeword on the $S$-th layer (cf. Fig. \ref{fig:Hier_Codebook_timecomp}). Each time when a user accesses the network, the beam search process introduced in Section III-A may be launched, and the cluster (angular grid) index for the particular user in discussion may be found. Hence, all the users are naturally grouped according to the angle grids, as shown in Fig. \ref{fig:BDMA_group}. Note that the user grouping is not fixed, as both the UAV and the ground users may move around. In practice, proper protocols need to be designed for the BS to manage the grouping information for all associated users.

The beamforming vectors for different users may be obtained based on the grouping information for all users. To illustrate this, suppose there are $U$ ($U\leq N_{\rm{RF}}$) users within different groups. After the grouping operation, the (BS, MS) beamforming codeword pairs for these $U$ users are found and represented by $\{({\bf{w}}_u$,${\bf{f}}_u)\}_{u=1,2,...,U}$, where ${\bf{w}}_u$ is the beam combining codewords for the $u$-th user at the BS side, ${\bf{f}}_u$ is the beamforming codewords for the $u$-th user at the MS side, each selected from their respective codebooks.  
Let ${\bf{H}}_u$ be the uplink channel response\footnote{With the per-user power constraint, the uplink transmission is dual to the downlink transmission, which means similar performance results can be observed for the downlink transmission.} for the $u$th user according to channel model Eq.~\eqref{eq_channel}.

With $\{{\bf{w}}_u\}_{u=1}^U$'s and $\{{\bf{f}}_u\}_{u=1}^U$'s as the receive and transmit beamforming vectors for all $U$ users, an effective $U\times U$ uplink channel ${\bf{H}}_{\rm{E}}$ may be obtained as ${\bf{H}}_{{\rm{E}}{i,j}}={\bf{w}}_i^{\rm{H}}{\bf{H}}_j{\bf{f}}_j$, and may be estimated at the receiver side (BS side). For such an equivalent $U\times U$ MIMO channel, minimum-mean-square-error (MMSE) detection with extra successive interference cancellation (SIC) receiver may be used \cite{TseFundaWC}.

\begin{figure}
\begin{minipage}[t]{0.5\linewidth}
\centering
\includegraphics[width=9.0 cm]{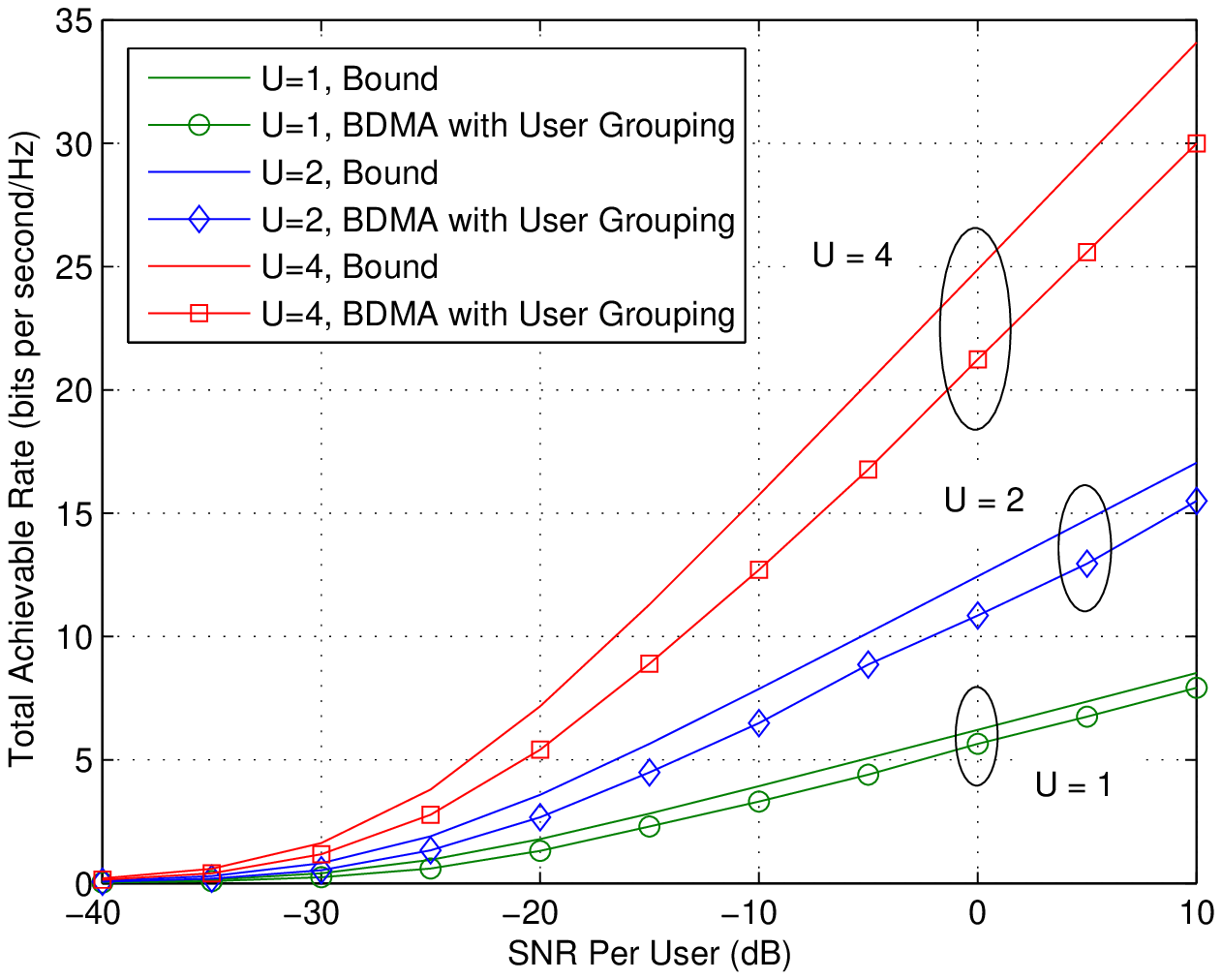}
\end{minipage}
\begin{minipage}[t]{0.5\linewidth}
\centering
\includegraphics[width=9 cm]{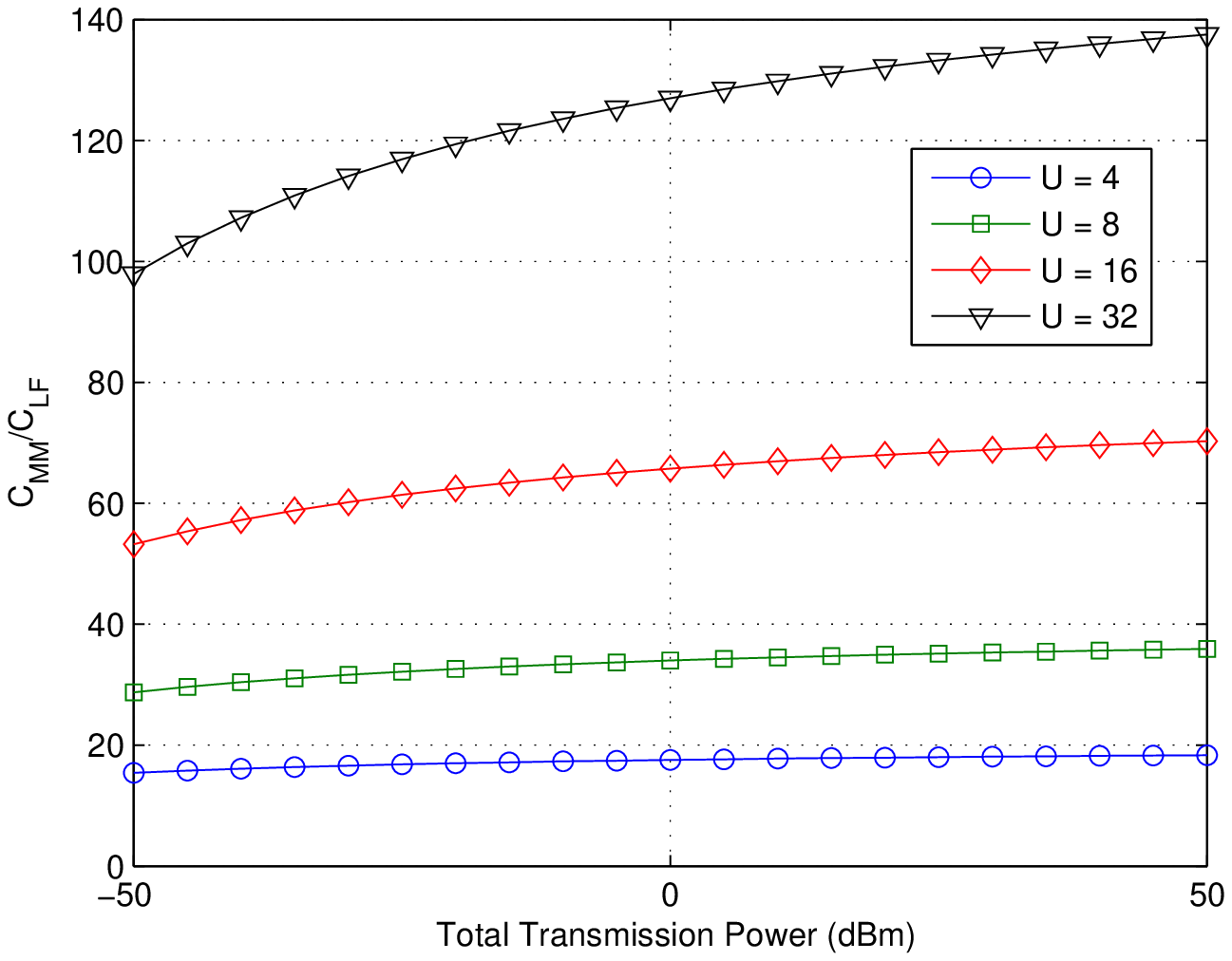}
\end{minipage}
\caption{\textbf{Left}: The total uplink achievable rate. Three MPCs are assumed with a LOS component and two 20 dB weaker NLOS components. \textbf{Right}: Comparison of multi-user capacity between the mmWave/low-frequency UAV cellular. The BS-MS distance is 1 km, and other parameters for the mmWave/low-frequency cellular are, respectively, 30 GHz/5 GHz for carrier frequency, 100 MHz/5 MHz for signal bandwidth, 24 dB/6 dB for BS array gain (256 versus 4 antennas), 12 dB/0 dB for MS array gain (16 versus 1 antennas).}
\label{fig:cmp_achie_rate_capacity}\vspace{-0.1 in}
\end{figure}

The left hand side of Fig. \ref{fig:cmp_achie_rate_capacity} shows the total achievable rate of the uplink transmission in the UAV cellular with mmWave SDMA/BDMA, where AoDs and AoAs for all users are estimated, user grouping is carried out using the BMW-SS codebook design, and an MMSE-SIC receiver is used at the BS side. The bound rate is computed, where AoDs and AoAs for all users are perfect, and the interference between the users are artificially forced to be zero. As we can see, slope of the mmWave SDMA performance curve is almost the same as that of the bound rate. This indicates that mmWave SDMA with proper user grouping is able to harness all the available degrees of freedom in the spatial domain. The performance loss relative to the bound rate is mainly due to the SNR loss.

An overall comparison of the multi-user capacity between the mmWave UAV cellular and the low frequency UAV cellular is shown on the right hand side of Fig.~\ref{fig:cmp_achie_rate_capacity}. The interference between the users are ignored for both cellulars just as the bound computation on the left hand side of Fig. \ref{fig:cmp_achie_rate_capacity}. The capacity of the mmWave UAV cellular is $C_{\rm{MM}}=UB_{\rm{MM}}\log_2(1+\rho_{\rm{MM}}/U)$, where $B_{\rm{MM}}$ is the signal bandwidth, $\rho_{\rm{MM}}$ is the received SNR of the LOS path incorporating both antenna gains and propagation loss, and $U$ is the number of users served by SDMA. The ergodic capacity of the low frequency UAV cellular is $C_{\rm{LF}}=\mathbb{E}\{4B_{\rm{LF}}\log_2(1+\rho_{\rm{LF}}|h|^2/4)$\}, where the factor 4 is the maximal number of users served by SDMA in LF due to the user area constraint (4 antennas at the BS in this figure), $B_{\rm{LF}}$ is the signal bandwidth, and $\rho_{\rm{LF}}$ is the averaged received SNR incorporating both antenna gains and propagation loss, $h$ is a standard complex Gaussian distributed variable to characterize the Rayleigh fading. From this figure we can find that the mmWave UAV cellular provides significantly higher multi-user capacity than the low-frequency UAV cellular, and the performance improvement mainly come from wider bandwidth and the capability of more SDMA users.



\section{Blockage}
Depending on the deploying environments, the probability that there exists a LOS link between the UAV BS and the ground MS, or the LOS probability, may vary. Typically when the UAV is deployed in rural areas, the LOS probability would be higher; and when the UAV is deployed in urban areas, the LOS probability would be lower due to potentially more blockage effect. Still the LOS probability of an air-to-ground link would be significantly higher than that of a ground-to-ground link due to the UAV elevation height. From a system point of view, an MS may be in one of the three following states depending on the LOS state, i.e., the LOS state during which a LOS link is available, the NLOS state during which a LOS link is broken while a lower rate communication link is still present (possibly thanks to reflection paths), and the outage state.

One of the major challenges for the mmWave UAV cellular, and for all mmWave communications in general, is the significant performance degradation in NLOS environments where the LOS path is blocked by obstacles, such as human bodies, buildings and others. Extensive measurement efforts have been carried out in \cite{roh2014millimeter} and it is reported that around 200 meter coverage is achievable for mmWave communications in NLOS environments despite the blockage effects. Similar observation is made in \cite{rangan2014millimeter}, where the presence of several distinct clusters of NLOS paths is reported. This is not entirely surprising since the adaptive beam training algorithms when properly designed has the ability to capture/track the strongest available paths, which are typically first-order and second-order reflection paths in NLOS environments. In \cite{zhao201328GHz}, measurements have shown that outdoor (where UAV mmWave communications typically occur) building materials are excellent reflectors, with the reflection coefficient as large as 0.896 for tinted glass.

Typically, the adaptive beam training algorithms may be designed with the possible loss of LOS path in mind and, other than the primary transmit/receive beam from the LOS path, maintain a short list of candidate transmit/receive beams, possibly from the first- and second-order reflection paths \cite{xiaozhenyu2013div}. Different signal processing techniques may be used to build the short list of candidate beams, such as power iteration, compressive sensing, and successive interference cancellation among others. Once the LOS path is lost, the short list of beams may be pursued instead to combat LOS blockage. Typically, choice of a lower modulation and coding (MCS) scheme rate is needed when one or more NLOS paths are used instead of the direct LOS path.

The blockage challenge is actually less severe for UAV air-to-ground mmWave communications, compared with regular ground mmWave communications. On one hand, because UAV is high above in the air, there is almost no reflection happening on the UAV side. In comparison, for regular mobile ground-to-ground mmWave communications, reflection happens on both the transmitter side and the receiver side because of their relatively low elevations. As a result, the overall reflection loss for UAV mmWave communications would be smaller.
More importantly, UAV enjoys fundamental capability of moving freely in the 3-dimensional space, subject to collision detection and avoidance. Hence, when the LOS path between the ground user and UAV is blocked, adaptive cruising algorithms may be developed to move the UAV to a new position such that a LOS path may be restored between UAV and the ground user. In comparison, such a LOS path restoration may be much more difficult, or even impossible, for regular mobile ground-to-ground mmWave communications. Moreover, since UAV is much easier to deploy, compared with the ground BS, multiple UAVs can provide additional diversities to combat with blockage, i.e., when the LOS link between a user and the UAV is blocked, the user can connect to another UAV where the LOS path is available.

\section{User Discovery}

For conventional wireless networks, a broadcast signal is periodically transmitted from the BS. Initially, the MS need to scan the available channels (e.g. physical broadcast channels PBCH) for broadcast signaling, which may include various system information such as regulatory information, network capability, and managing information. Before an MS is allowed to transmit data to the MS, a random access procedure need to be carried out. In particular, the MS may transmit a random access preamble to the target BS, and the BS would respond with a random access response, which would include various system and control information, such as timing advance adjustment, uplink grant etc. With the uplink grant, the MS may proceed to transmit a radio resource control (RRC) connection request, and the BS may respond with a RRC connection setup response including the radio resource configuration information.

Such a user discovery would not work well directly for mmWave UAV communications. The reason is that, to overcome significant path loss and improve the communication range, mmWave communications entail inherently directional transmissions at least from the UAV side. Because of the directional transmission/reception, the UAV may not be able to hear the initial random access preamble from the MS, or the MS may not be able to hear the broadcast signals from the UAV. One possible solution is to let the UAV BS transmit multiple directional broadcast signals in multiple directions over different time slots to mimic an omni-directional antenna pattern. Thus, a PBCH-scanning MS would detect at least one directional transmission of the broadcast signal, and may proceed to send a random access preamble afterwards in a proper time slot, during which the UAV may operate in a proper directional receiving mode. The BS may respond with a directional random access response, including possibly timing advance adjustment, uplink grant as well as antenna sector information of the MS (note that MS itself may transmit/receive directionally). RRC connection request and response may follow similarly. Drawback of the above process is that multiple directional transmissions are necessary to perform user discovery, and hence  non-negligible overhead. To keep the user discovery overhead under control, very fine beam training is not recommended during the user discovery stage. Instead, coarse beam training may be used at this stage, while a finer beam training may be performed afterwards, before payload transmissions.

\begin{figure}[t]
\begin{center}
  \includegraphics[width=12 cm]{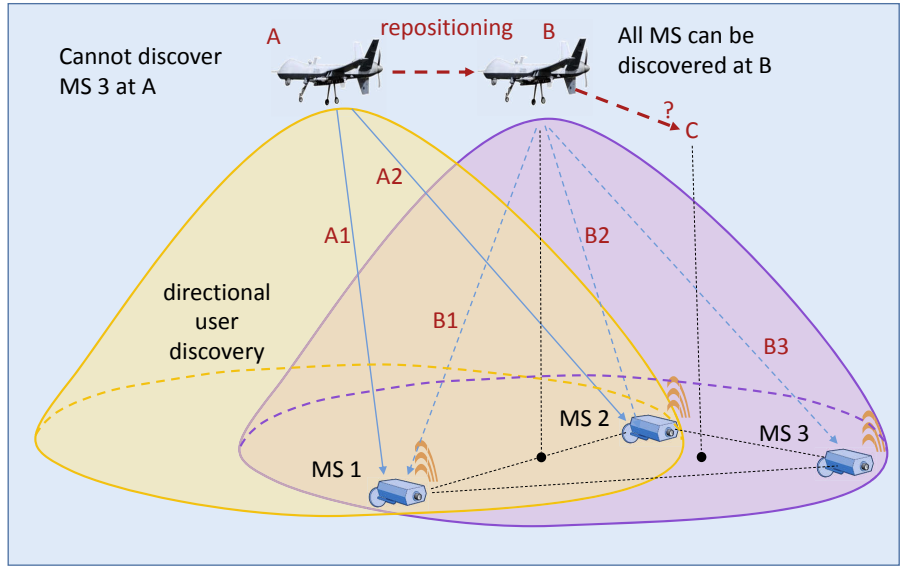}
  \caption{User discovery in a mmWave UAV cellular with re-positioning.}
  \label{fig:userdiscovery}\vspace{-0.1 in}
\end{center}
\end{figure}

Unique to UAV mmWave communications is that the UAV positioning intertwines with the user discovery. As illustrated in Fig. \ref{fig:userdiscovery}, the UAV BS initially is positioned at point A and is able to discover MS 1 and 2, at direction A1 and direction A2 respectively. Once the user discovery of MS 1 and 2 is completed, it may be worthwhile for the BS to re-position to point B, midpoint of MS 1 and 2, to further improve the network performance. Whether this should be done is a tradeoff between the network performance improvement due to repositioning and the associated signaling cost to maintain the network after repositioning. If  BS indeed decides to reposition to point B after weighing the tradeoff, then BS need to update the direction of MS 1 and 2 to direction B1 and direction B2 respectively. How to achieve this efficiently is yet to be defined. Furthermore when the UAV repositions to point B, it may discover a new MS 3 which was not found at point A due to the limited range. At this point, the UAV needs to weigh again the tradeoff between the performance improvement moving from point B to point C (centroid point of MS 1, 2 and 3) and the associated signaling cost to maintain the network of MS 1, 2 and 3. In general, the UAV positioning and directional mmWave user discovery may be carried out in an iterative manner.

\section{Conclusions}

To support high data rate urgent or ad hoc communications, we consider mmWave UAV cellular networks and the associated challenges and solutions. In particular, hierarchical beamforming codebook structure is investigated as an enabling method for fast beamforming training and tracking. Numerical results demonstrate that the BMW-SS codebook design is able to generate beams with different widths, corresponds to beams on different layers, and achieves excellent beam detection performance. It also scales nicely for very large antenna arrays. Although the overall channel itself may experience fast Doppler due to UAV movement, the major multipath components are shown to undergo only slow variation thanks to high gain directional transmissions. Millimeter wave SDMA is also investigated, where directional user grouping may be used to classify users into different spatial groups, and only user from different groups may access the BS at the same time using SDMA. Significant capacity improvement is possible and mainly due to the large signal bandwidth and the use of SDMA in the spatial domain.

The blockage problem, a serious performance limiting factor for regular mmWave cellular networks, may actually be alleviated thanks to UAV movement. Intelligent cruising algorithms need be developed to enable UAV to fly out of a blockage zone and reestablish a LOS link to the MS. Finally, the relationship of UAV positioning and mmWave directional user discovery is studied. On one hand, with a fixed position, UAV would be able to discover only the nearby users. On the other hand, UAV needs to find all potential users to serve to optimize its self positioning. In general, the mmWave directional user discovery and UAV positioning may be carried out in an iterative manner to keep improving the network performance.

\section*{Acknowledgments}

This work was partially supported by the National Natural Science Foundation of China (NSFC) under grant Nos. 61571025, 91338106, 91538204, and 61231013, National Basic Research Program of China under grant No.2011CB707000, and Foundation for Innovative Research Groups of the National Natural Science Foundation of China under grant No. 61221061.


\section*{Biographies}

\textsc{Zhenyu Xiao} (xiaozy@buaa.edu.cn) is a researcher in Beihang University, Beijing, China. He has published over 50 papers, and served as reviewers for many famous journals. He has been TPC members of IEEE GLOBECOM'12, IEEE WCSP'12, IEEE ICC'15, etc. His research interests are millimeter-wave communications and UAV networks.

\textsc{Pengfei Xia} (pengfei.xia@gmail.com) is a Full Chair Professor with Tongji University, Shanghai, China. His research interests are wireless communications, networks, and signal processing. He is a coeditor of the book 60GHz Technology for Gbps WLAN and WPAN: From Theory to Practice. Currently, he serves as SPCOM technical committee member and SPCOM Industrial/Government Subcommittee Chair for the IEEE Signal Processing Society. He was the recipient of the IEEE Signal Processing Society Best Paper Award 2011.

\textsc{Xiang-Gen Xia} [M¡¯97, SM¡¯00, F¡¯09] (xxia@ee.udel.edu) received his Ph.D. in electrical engineering from University of Southern California, Los Angeles, in 1992. He is currently the Charles Black Evans Professor in the Department of Electrical and Computer Engineering, University of Delaware, Newark, Delaware, USA. His current research interests include space-time coding, MIMO and OFDM systems, digital signal processing, and SAR and ISAR imaging. He is the author of the book Modulated Coding for Intersymbol
Interference Channels (New York, Marcel Dekker, 2000).

\end{document}